\documentclass{article}
\usepackage{spconf,graphicx}
\usepackage{epsfig}
\usepackage{amsmath,amssymb,amsthm,amsfonts}
\usepackage{hyperref}

\newcommand{\bqn}{\begin{eqnarray}}
	\newcommand{\eqn}{\end{eqnarray}}
\newcommand{\bq}{\begin{eqnarray*}}
	\newcommand{\eq}{\end{eqnarray*}}

\newcommand{\ba}{\left( \begin{array}}
\newcommand{\ea}{\end{array} \right)}

\usepackage{xcolor}

\graphicspath{{figures/}}

\title{Causality as a Minimum Energy Principle}
\name{Moo K. Chung$^1$
\thanks{Published in IEEE Engineering in Medicine and Biology Society Annual Conference (EMBC) 2026. 
The correspondence should be sent to {\tt mkchung@wisc.edu}.}
, D. Vijay Anand$^2$, 
Anass B El-Yaagoubi$^3$, Jae-Hun Jung, Anqi Qiu, Hernando Ombao}
\address{$^1$University of Wisconsin, Madison, USA., 
$^2$University College London, UK,\\
$^3$POSTECH, Korea, $^4$Hong Kong Polytechnic University, China, 
$^5$KAUST, Saudi Arabia}

\begin{document}
\maketitle

\begin{abstract}
Classical causal models, such as Granger causality and structural equation modeling, are largely restricted to acyclic interactions and struggle to represent cyclic and higher-order dynamics in complex networks. We introduce a causal framework grounded in a variational principle, interpreting causality as directional energy flow from high- to low-energy states along network connections. Using Hodge theory, network flows are decomposed into dissipative components and a persistent harmonic component that captures stable cyclic interactions. Applied to resting-state fMRI connectivity, our variational framework reveals robust cyclic causal patterns that are not detected by conventional causal models, highlighting the value of variational principles for causality.
\end{abstract}


\section{Introduction}

Understanding causality is fundamental to scientific inquiry, as it enables the identification of mechanisms that govern system dynamics beyond mere statistical associations. In neuroscience, causal modeling is essential for elucidating how brain regions influence one another and how information propagates through neural circuits to support cognition and behavior \cite{bourakna.2024,lee.2014.MICCAI}. Unlike correlation-based approaches \cite{anand.2024.ISBI}, which capture symmetric dependencies, causal models aim to infer asymmetric, directional relationships.

Traditional causal methods, including Granger causality and structural equation modeling, have been widely applied to neuroimaging data but typically rely on assumptions of {\it acyclicity} and linearity \cite{geweke.1982,ombao.2024}. These constraints limit their ability to represent recurrent and higher-order interactions that are ubiquitous in brain networks. Moreover, most functional connectivity models emphasize pairwise relationships between regions, overlooking higher-order dependencies that are increasingly recognized as central to brain network organization \cite{battiston.2020,sizemore.2019}. While the relevance of such interactions is increasingly acknowledged, progress has been impeded by the lack of scalable analytical tools modeling them.

To move beyond conventional acyclic causal models, we introduce a foundational variational causal framework grounded in the minimum energy principle. This formulation defines causal structure as the solution to an energy-minimizing flow problem on a network, allowing recurrent and multiscale dynamics to arise naturally from the variational objective rather than being imposed a priori. The framework leverages \emph{Hodge theory}, a mathematical formalism that unifies simplicial topology with spectral geometry \cite{anand.2023.TMI,lee.2014.MICCAI}, to uniquely partition network edge flows into orthogonal dissipative and persistent components \cite{anand.2024.ISBI}. Within this decomposition, harmonic flows emerge as persistent cyclic interactions that cannot be eliminated without increasing the system energy, providing a principled representation of irreducible recurrent structure. This variational separation yields a topology-aware and noise-robust dissection of functional connectivity that is explicitly sensitive to higher-order and cyclic interactions often missed by classical causal models. The study is expected to offer, for the first time, a mathematically rigorous and topologically grounded variational theory of causal structure that elevates causality from a model-specific assumption to a first-principles consequence of energy minimization on networks.

We apply the proposed framework to resting-state functional magnetic resonance imaging (fMRI) data from human subjects to demonstrate its practical utility in uncovering cyclic causal organization in the human brain. Our results reveal a substantial and reproducible harmonic component that is stable across subjects and time, yet is not captured by conventional causal models, thereby providing new insight into long-range recurrent interactions underlying resting-state dynamics. These findings illustrate how variational and topological principles together offer a complementary, mechanistically interpretable foundation for causal inference in complex brain networks.

\section{Materials and Methods}

\subsection{Imaging Data and Preprocessing}

We analyzed rs-fMRI data from 400 healthy adults (168 males, 232 females; age $22$–$36$ years, mean $29.24 \pm 3.39$)  from the Human Connectome Project (HCP). Data were acquired on Siemens 3T Connectome Skyra scanners, with acquisition parameters and the HCP minimal preprocessing pipeline described in \cite{glasser.2016}. Volumes with framewise displacement exceeding $0.5$ mm, together with adjacent frames, were scrubbed, and subjects with excessive motion were excluded \cite{huang.2020.NM}. The Automated Anatomical Labeling (AAL) atlas \cite{tzourio.2002} was used to parcellate each brain into 116  regions. Regional rs-fMRI signals were obtained by averaging voxel time series within each region, yielding 116 time series of length $1200$ per subject \cite{huang.2020.NM}. 

\subsection{Simplicial Complexes from rs-fMRI Data}

Using a fixed temporal lag \(l=20\) seconds, we compute time-lagged Pearson correlations within 40-second sliding windows,
\(
\rho_{ij}(t)=\mathrm{corr}\!\big(Y_i(t),\,Y_j(t+l)\big),
\)
which yield a time-varying asymmetric connectivity matrix \(\rho(t)\). The chosen lag is substantially longer than typical hemodynamic delays and is intended to probe delayed, large-scale interactions while reducing sensitivity to region-specific vascular timing differences \cite{keilholz.2017}. For each unordered pair \((i,j)\), a dominant temporal interaction is identified by comparing \(\rho_{ij}(t)\) and \(\rho_{ji}(t)\). The larger value is retained as the forward direction, and directionality is encoded through an antisymmetric edge-flow matrix \(X(t)=(X_{ij}(t))\), defined such that \(X_{ij}(t)>0\) represents dominant influence from region \(i\) to region \(j\), while \(X_{ji}(t)=-X_{ij}(t)\).

To suppress weak or spurious connections, edge flows are thresholded at correlation \(\tau=0.6\); if both directions fall below the threshold, the pair is discarded. The surviving edges define the $1$--skeleton of a simplicial complex. A triangle \((i,j,k)\) is included when all three corresponding edge magnitudes exceed the threshold and their orientations are consistent, either all clockwise or all counterclockwise, indicating a coherent cyclic interaction. Higher-dimensional simplices may be defined analogously.

\subsection{Boundary and Coboundary Operators}

The boundary operator generalizes the incidence matrix by describing how each \(p\)-simplex is composed of its \((p\!-\!1)\)-faces. This relationship is represented by a signed incidence matrix \(\mathbf{B}_p\), whose entries are \(+1\), \(-1\), or \(0\) depending on relative orientation \cite{anand.2024.ISBI,huang.2025}. \(\mathbf{B}_1 \in \mathbb{R}^{|V|\times|E|}\) encodes the node--edge incidence while \(\mathbf{B}_2 \in \mathbb{R}^{|E|\times|F|}\) encodes the edge--face incidence. The coboundary operator is given by \(\mathbf{B}_{p+1}^\top\); for example, \(\mathbf{B}_1^\top\) maps node potentials to edgewise gradients, while \(\mathbf{B}_2^\top\) maps edge flows to curls around triangles.

Let \(\mathcal{C}^k\) denote the space of functions on \(k\)-simplices, including vertex potentials (\(\mathcal{C}^0\)), edge flows (\(\mathcal{C}^1\)), and triangle circulations (\(\mathcal{C}^2\)). The boundary operator \(\mathbf{B}_1:\mathcal{C}^1\to\mathcal{C}^0\) maps an edge flow to its net incidence at each vertex. Its kernel,
\[
\ker(\mathbf{B}_1)=\{X\in\mathcal{C}^1:\mathbf{B}_1X=0\},
\]
consists of closed edge flows, or \(1\)-cycles. These cycles are related to the image of \(\mathbf{B}_2:\mathcal{C}^2\to\mathcal{C}^1\), which maps oriented triangles to their boundary edges. Consequently,
\(
\mathbf{B}_1\mathbf{B}_2=0,
\)
i.e., the boundary of a boundary is empty \cite{anand.2023.TMI}.

\subsection{Variational Principle for Causality}

An edge flow \(X \in \mathcal{C}^1\) can be interpreted as fluid-like transport or information flow on a network. From a variational perspective, causality corresponds to the tendency of such flows to evolve toward the configurations of minimum energy state. We define  \emph{1-Hodge Laplacian}
\(
\Delta_1=\mathbf{B}_1^\top\mathbf{B}_1+\mathbf{B}_2\mathbf{B}_2^\top,
\)
along edges \cite{anand.2024.ISBI,anand.2023.TMI}.  The associated Dirichlet energy
\[
E(X)=\frac{1}{2}\langle X,\Delta_1 X\rangle
=\frac{1}{2}\|\mathbf{B}_1 X\|_2^2+\frac{1}{2}\|\mathbf{B}_2^\top X\|_2^2,
\]
quantifies nodal divergence through \(\mathbf{B}_1 X\) and local circulation through \(\mathbf{B}_2^\top X\). Causal propagation is viewed as the evolution of edge flows toward energy-minimizing states.

Gradient descent on \(E(X)\) yields the Dirichlet diffusion
\begin{equation}
\frac{dX(t)}{dt}=-\Delta_1 X(t),
\label{eq:diffusion}
\end{equation}
which evolves \(X(t)\) in the direction of steepest energy descent \cite{noirhomme.2024}.
Along this trajectory, the Dirichlet energy decreases monotonically
\(
\frac{d}{dt}E(X(t))
=-\|\Delta_1 X(t)\|_2^2 \le 0,
\)
inducing dissipative dynamics. To account for ongoing spontaneous fluctuations in rs-fMRI, we extend this formulation to a driven--dissipative system \cite{breakspear.2017}:
\(
\frac{dX(t)}{dt} = -\Delta_1 X(t) + U(t),
\)
where \(U(t)\) represents latent endogenous neural activity driving these fluctuations. 
Under this dynamics, the edge flow admits the orthogonal decomposition \cite{anand.2024.ISBI,chung.2026.ISBI}
\bqn
X(t)=X_D(t)+X_H(t), \label{eq:HD}
\eqn
where \(X_D(t)\) is attenuated over time, while \(X_H(t)\) lies in the cycle space and encodes stable recurrent organization. This yields a topology-aware causal filter that removes unstable fluctuations while retaining coherent low-energy structure. This induces a topology-aware filtering of causal interactions that suppresses unstable fluctuations while retaining structurally stable patterns. The approach parallels invariance-based causal inference \cite{peters.2016}, where causality is defined through stability under perturbations. Our framework provides a data-driven variational representation of causal interactions rather than a physiological model.

\subsection{Numerical Implementation}
Computation is performed within each sliding window, where \(U(t)\) is assumed locally constant and thus $X_H$ is not time-varying. The decomposition (\ref{eq:HD}) is 
a special case of the Hodge decomposition \cite{anand.2024.ISBI}\footnote{An extensive MATLAB library for various Hodge computations is available at \url{https://github.com/laplcebeltrami/hodge}.}. The harmonic component satisfies
\[
X_H \in \ker(\mathbf{B}_1)\cap\ker(\mathbf{B}_2^\top)
\]
and obtained via orthogonal projection
\[
X_H = \arg\min_{Z}\|X-Z\|^2 \;\text{subject to}\; \mathbf{B}_1 Z=0,\;\mathbf{B}_2^\top Z=0.
\]
Let
$B=\begin{bmatrix}\mathbf{B}_1^\top & \mathbf{B}_2\end{bmatrix}^\top$ and \(
\mathcal{P}_H=I-B^\top(BB^\top)^\dagger B.
\)
Then
\[
X_H=\mathcal{P}_H X, \qquad X_D=X-X_H.
\]

\begin{figure}[t]
	\centering
	\includegraphics[width=1.0\linewidth]{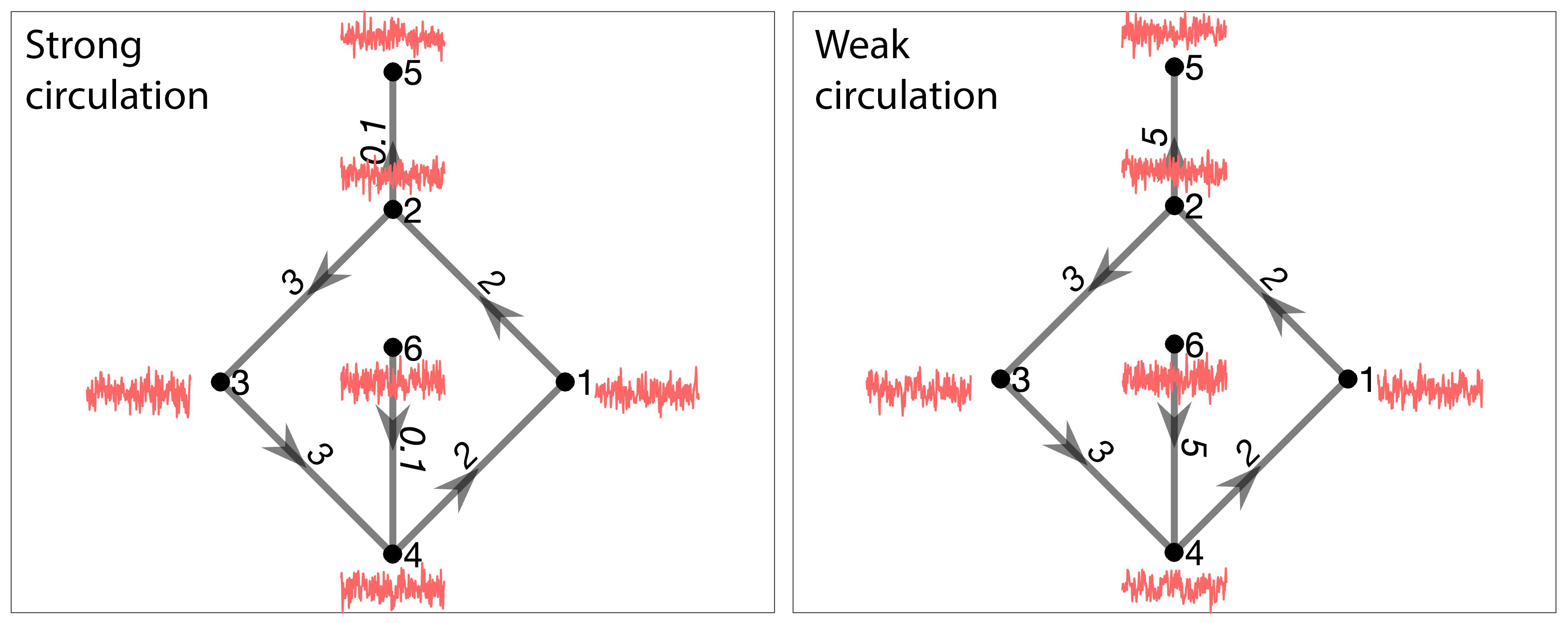}
	\caption{\footnotesize Six-node network with a square-shaped 1-cycle. 
Left: small edge flows ($2 \to 5$ and $6 \to 4$ set to 0.1) strengthening the circulation along the 1-cycle. 
Right: large edge flows ($2 \to 5$ and $6 \to 4$ set to 5) weaken the circulation along the 1-cycle. Time series at each node is generated by the VAR-process following the underlying ground-truth edge flow. }
\label{fig:simulation-circulation}
\end{figure}

\section{Validation}

Most widely used causal models are built on pairwise interactions and acyclic graph structures. Methods such as Granger causality, structural equation models (SEMs), and many DAG-based approaches posit explicitly on a directed acyclic graph, thereby precluding feedback and circulation by design \cite{geweke.1982,ombao.2024}. Consequently, cyclic structure is often ruled out or treated as a model violation rather than a target of inference and learning. In this simulation study\footnote{Extensive MATLAB simulation codes with detailed documentation, including additional comparisons not reported in the paper, are available at \url{https://github.com/laplcebeltrami/harmonic}.}, we contrast the proposed causal framework with existing methods.

{\it Ground Truth.} We constructed a ground-truth directed graph with six nodes containing a single square-shaped $1$-cycle (Fig.~\ref{fig:simulation-circulation}). Two baseline configurations were considered, corresponding to strong and weak circulation along the four-node cycle ($1 \to 2 \to 3 \to 4 \to 1$). Time series were generated using a vector autoregressive (VAR) process of order $L=10$, with coefficient matrices chosen to be consistent with the prescribed directed cycle,
\(
Z_t=\sum_{\ell=1}^L A_\ell Z_{t-\ell}+\varepsilon_t,  
\varepsilon_t\sim\mathcal{N}(0,\sigma^2 I),
\)
using $200$ time points per simulation. The use of longer lags progressively obscures the explicit cycle structure, making causal recovery more challenging.

\begin{table}[t!]
\centering
\caption{\footnotesize Cosine similarity (mean $\pm$ std) over 100 trials for strong and weak circulation along a 1-cycle (Fig.~\ref{fig:simulation-circulation}) across different noise levels $\sigma$.}
\label{table:cosine}
\setlength{\tabcolsep}{4pt}
\scriptsize
\begin{tabular}{c|c|c|c|c|c}
\hline
Circ. & Noise  & Granger & SEM & Corr. & {\bf Harmonic} \\
\hline
Strong
 & 0.1  & $0.34 \pm 0.09$ & $0.38 \pm 0.05$ & $0.84 \pm 0.05$ & $\mathbf{0.97 \pm 0.01}$ \\
 & 1    & $0.33 \pm 0.09$ & $0.38 \pm 0.04$ & $0.83 \pm 0.05$ & $\mathbf{0.96 \pm 0.03}$ \\
 & 10   & $0.34 \pm 0.10$ & $0.38 \pm 0.04$ & $0.83 \pm 0.05$ & $\mathbf{0.98 \pm 0.01}$ \\
\hline
Weak
 & 0.1  & $0.30 \pm 0.09$ & $0.38 \pm 0.04$ & $0.58 \pm 0.05$ & $\mathbf{0.63 \pm 0.07}$ \\
 & 1    & $0.30 \pm 0.07$ & $0.38 \pm 0.04$ & $0.60 \pm 0.04$ & $\mathbf{0.65 \pm 0.06}$ \\
 & 10   & $0.31 \pm 0.08$ & $0.37 \pm 0.04$ & $0.59 \pm 0.05$ & $\mathbf{0.63 \pm 0.09}$ \\
\hline
\end{tabular}
\end{table}

{\it Performance metric.} Cosine similarity \cite{luo.2018} is used to compare the estimated edge flow with the ground-truth flow along the $1$-cycle. As a scale-invariant measure, it evaluates alignment of orientation independently of magnitude. This is appropriate since the harmonic component encodes cyclic structure primarily through directional consistency of the flow.

{\it Results.} Granger causality was estimated using a VAR model with lag order selected by minimizing prediction error \cite{geweke.1982,ombao.2024}, and edge flows were defined as $1-p_{ij}$ from $F$-tests. Despite being aligned with the VAR-based data generation, Granger causality performs poorly in the presence of cyclic structure.  SEM was implemented using nonlinear additive regression, with each node predicted from all others and edge weights defined by permutation importance. While capturing nonlinear dependencies, it shows limited ability to recover cyclic structure \cite{buhlmann.2014}. Correlation-based directionality, used as input to harmonic flow, yields moderate recovery for strong circulation but degrades substantially for weak cycles.

In contrast, the proposed harmonic flow method achieves consistently superior recovery of the ground-truth cycle. It attains near-perfect alignment for strong circulation and remains clearly superior for weak circulation, outperforming Granger causality, SEM, and correlation. Performance is stable across noise levels, including high noise ($\sigma=10$), demonstrating robustness in isolating persistent cyclic structure.

\section{Results}

\subsection{Emergence of Stable Cyclic Structures in rs-fMRI}

\begin{figure}[t!]
\centering
\includegraphics[width=1\linewidth]{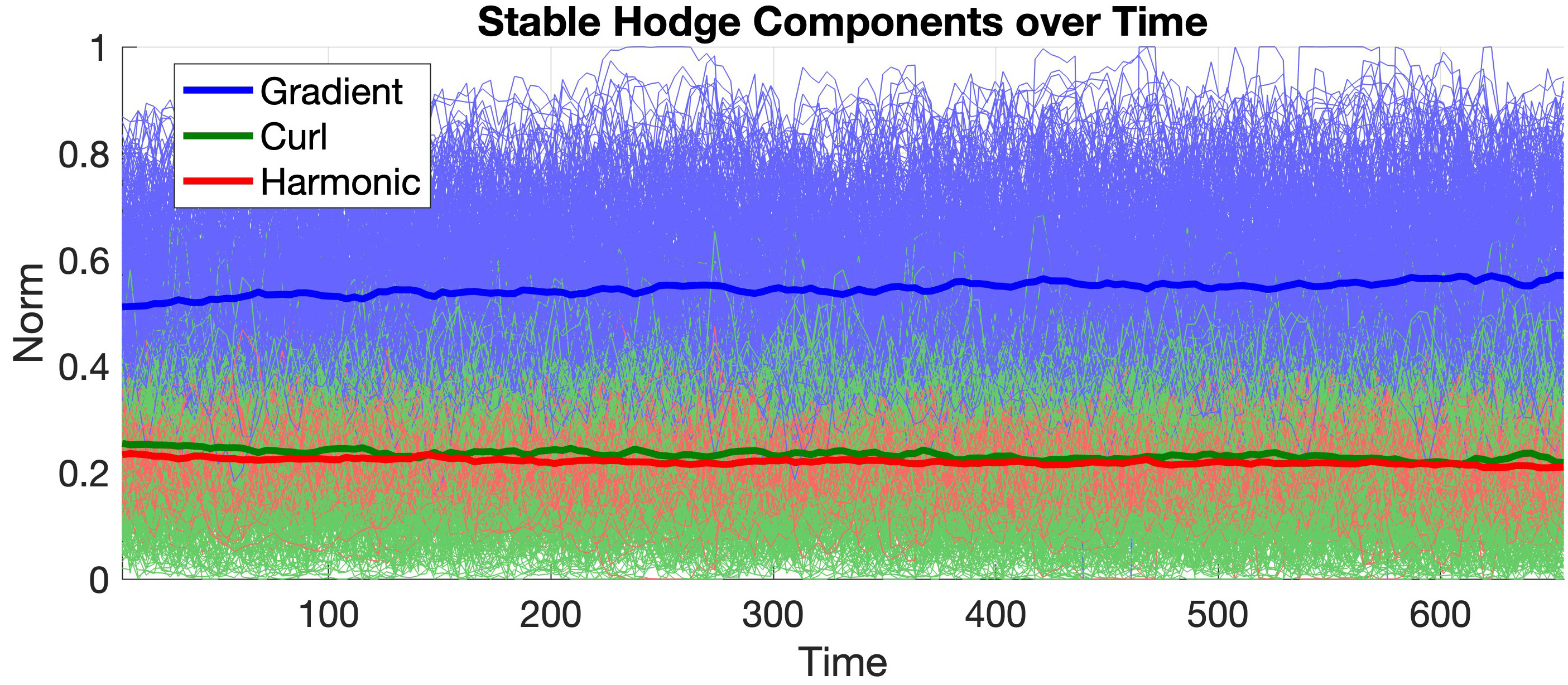}
\caption{\footnotesize Temporal evolution of harmonic flow in dynamic resting-state fMRI networks. Thin lines show individual trajectories from 400 subjects, and the thick line indicates the across-subject mean. The dissipative component \(X_D\) aggregates non-harmonic (gradient and curl) contributions.}
\label{fig:loopanalysis}
\end{figure}

We discarded fMRI frames before 72 seconds and after 720 seconds, and estimated Hodge flow at every 3.6-second interval (5 TRs). For each time window, the edge flow \(X\) was decomposed into dissipative \(X_D\) and harmonic \(X_H\) components. Across 400 subjects and all time frames, the harmonic component accounted for 21.97$\pm$6.16$\%$ of total flow energy ($p=0.004$), computed as \(\|X_H\|_2^2 / \|X\|_2^2\) averaged over time and subjects, indicating stable cyclic organization (Fig.~\ref{fig:loopanalysis}). While \(X_D\) dominated overall dynamics, \(X_H\) consistently captured long-range cycles that persisted across time and subjects.

Both components remained stable over the scan, with modest temporal redistribution of energy consistent with prior report of strengthening large-scale correlations during prolonged rest \cite{birn.2013}. This stability reflects the low-dimensional, topologically constrained nature of the harmonic subspace, which suppresses noise while allowing gradual evolution under the sliding-window estimation. Together, these results provide direct evidence that persistent cyclic interactions form a fundamental backbone of resting-state brain dynamics.

\begin{figure}[t!]
\centering
\includegraphics[width=1\linewidth]{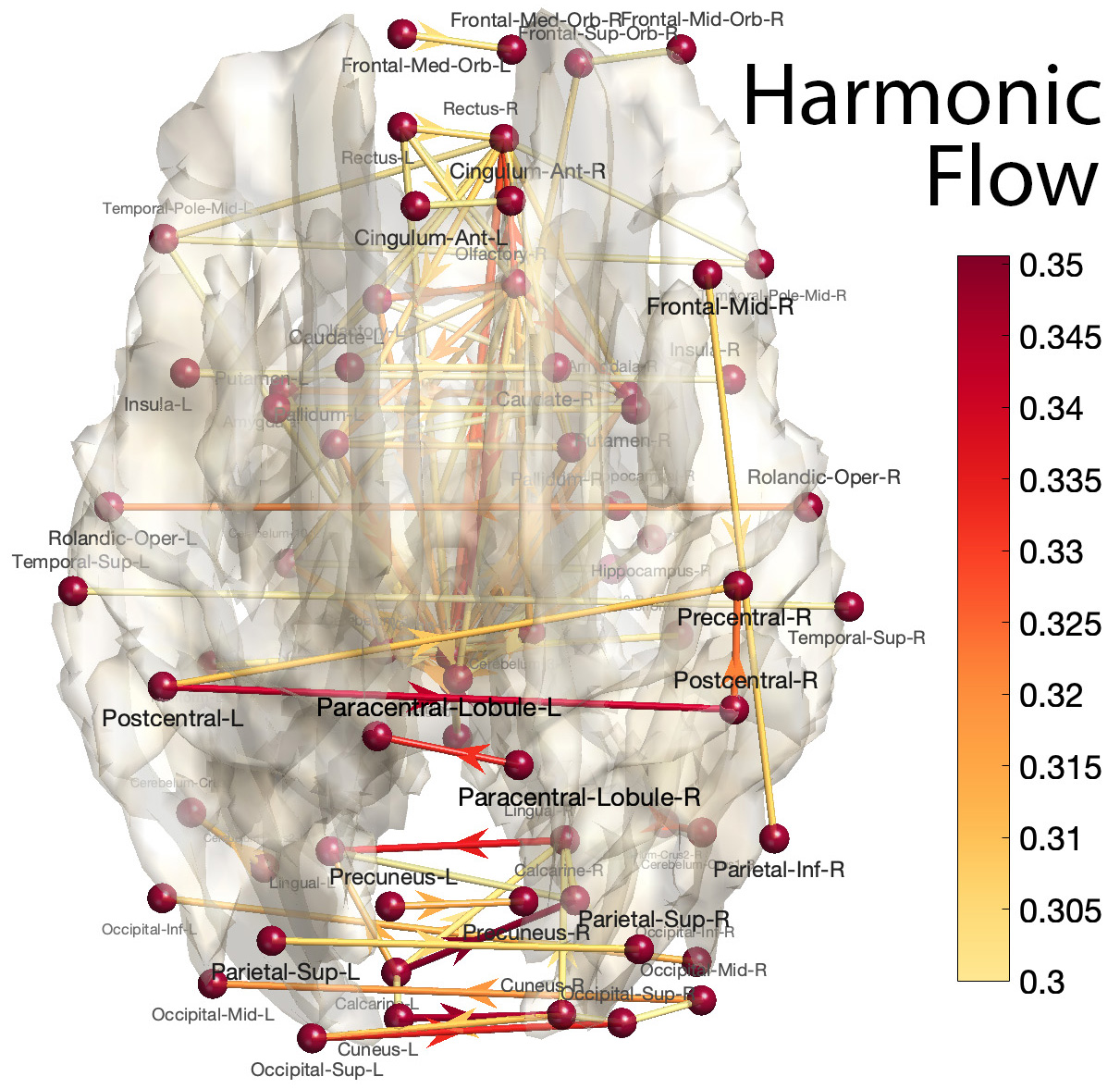}
\caption{\footnotesize Average harmonic flows across time and 400 subjects, thresholded at  0.3. The strongest harmonic flows predominantly correspond to interhemispheric interactions, particularly across homotopic regions in the primary sensory and motor cortices, where signals circulate bidirectionally between the left and right hemispheres.}
\label{fig:averageharmonic}
\end{figure}

\subsection{Identification of Stable Cyclic Causality in rs-fMRI}

Harmonic components, representing equilibrium states of brain network dynamics, capture dominant cyclic patterns that have been largely overlooked in prior work. We report the top 10 dominant harmonic edges. To characterize cohort-level structure, harmonic edge flows \(X_H(t)\) were averaged over time within each subject and then across subjects (Fig.~\ref{fig:averageharmonic}).

Two dominant organizational motifs emerge. First, the strongest harmonic flows cluster in interhemispheric homotopic pairs, including regions such as the calcarine cortex, cuneus, postcentral gyrus, lingual gyrus, superior occipital cortex, and paracentral lobule (Table~\ref{table:20harmonic}). These homotopic connections form stable cyclic backbones rather than transient feedforward interactions, reflecting recurrent left--right dynamics. Second, prominent harmonic flows involve the vermis and adjacent cerebellar lobules together with olfactory and limbic regions, forming stable cerebello--limbic loops that indicate persistent subcortical--cortical feedback structure during rest.

Overall, dominant harmonic flows exhibit a strongly bilateral organization, with nearly half originating within or between the left and right hemispheres and more than \(43\%\) corresponding to cross-hemispheric interactions. Midline structures contribute a smaller fraction (approximately \(7\%\)) but show strong bilateral connectivity, suggesting a modulatory role in coordinating recurrent cortical and subcortical circuits. These findings indicate that stable cyclic causality forms a large-scale recurrent backbone spanning bilateral and midline systems in the resting brain.

\begin{table}[t!]
\centering
\caption{Top 10 dominant harmonic flows across brain regions. Interhemispheric homotopic pairing constitutes the majority of the strongest flows.}
\label{table:20harmonic}
\scriptsize
\begin{tabular}{r l c l c}
\hline
Rank & From &  & To & Magnitude \\
\hline
 1 & Calcarine-L       & $\to$ & Calcarine-R          & 0.3635 \\
 2 & Vermis-10         & $\to$ & Vermis-1-2           & 0.3632 \\
 3 & Cuneus-L          & $\to$ & Cuneus-R             & 0.3608 \\
 4 & Postcentral-L     & $\to$ & Postcentral-R        & 0.3559 \\
 5 & Olfactory-R       & $\to$ & Vermis-10            & 0.3509 \\
 6 & Lingual-R         & $\to$ & Lingual-L            & 0.3460 \\
 7 & Occipital-Sup-L   & $\to$ & Occipital-Sup-R      & 0.3440 \\
 8 & Paracentral-Lobule-R & $\to$ & Paracentral-Lobule-L & 0.3430 \\
 9 & Olfactory-R       & $\to$ & Vermis-1-2           & 0.3425 \\
10 & Vermis-10         & $\to$ & Rectus-R             & 0.3402 \\
\hline
\end{tabular}
\end{table}

\section{Discussion}
A substantial harmonic component (22\%) of resting-state activity reflects long-range cyclic organization, with direct implications for brain network topology and dynamics. Since harmonic energy arises only in the presence of noncontractible loops, quantified by a positive first Betti number \(\beta_1\), it captures genuine cyclic structure beyond tree-like propagation. The harmonic subspace \(\ker(\mathbf{B}_1)\cap\ker(\mathbf{B}_2^\top)\) thus represents an intrinsic topological motif.

Harmonic flow serves as a stable backbone of resting-state dynamics. While dissipative components reflect transient propagation, the harmonic component encodes persistent circulation patterns, prominently observed in interhemispheric and cerebellar–limbic loops, indicating large-scale recurrent coupling. This stability follows from its confinement to a low-dimensional subspace with dimension \(\beta_1 \ll |E|\) \cite{chung.2026.ISBI}, where \(\beta_1\) is the first Betti number (the dimension of the cycle space), making it robust to temporal fluctuations and noise. Thus, time-varying dynamics can be captured by estimating harmonic flow within sliding windows, allowing gradual evolution while preserving the underlying topology.

\bibliographystyle{IEEEbib}

\end{document}